# Real-time imaging of atomic potentials in 2D materials with 30 keV electrons

*Sytze de Graaf*[1,*], *Majid Ahmadi*[1], *Ivan Lazić*[2], *Eric G.T. Bosch*[2], *Bart J. Kooi*[1,*]

[1]Zernike Institute for Advanced Materials, University of Groningen, Nijenborgh 4, 9747 AG Groningen, Netherlands.

[2]Thermo Fisher Scientific, Achtseweg Noord 5, 5651 GG Eindhoven, Netherlands.

Scanning transmission electron microscopy (STEM) is the most widespread adopted tool for atomic scale characterization of two-dimensional (2D) materials. Many 2D materials remain susceptible to electron beam damage, despite the standardized practice to reduce the beam energy from 200 keV to 80 or 60 keV. Although, all elements present can be detected by atomic electrostatic potential imaging using integrated differential phase contrast (iDPC) STEM or electron ptychography, capturing dynamics with atomic resolution and enhanced sensitivity has remained a challenge. Here, by using iDPC-STEM, we capture defect dynamics in 2D $WS_2$ by atomic electrostatic potential imaging with a beam energy of only 30 keV. The direct imaging of atomic electrostatic potentials with high framerate reveals the presence and motion of single atoms near defects and edges in $WS_2$ that are otherwise invisible with conventional annular dark-field STEM or cannot be captured sufficiently fast by electron ptychography.

2D materials are, and have been for the last decade, a hot topic of research for their large potential in future applications by virtue of their reduced dimensions and tunable properties. Also, novel properties can emerge that are not present in the bulk. This emerging platform of materials has found its use in various fields such as spintronics[1], neuromorphics[2], optoelectronics[3], sensors[4], catalysis[5] or DNA sequencing[6].

Scanning transmission electron microscopy (STEM) is the most widespread adopted tool for direct observation of crystalline structure, grain boundaries and defects in 2D materials. The main advantage of scanning an electron probe (with STEM) over parallel electron wave illumination (with TEM), is the versatility to employ multiple electron detectors for imaging as well as for spectroscopy purposes, such that simultaneous atomically resolved images and chemical maps can be acquired. The interpretation of atomic resolution images is generally also more straightforward for STEM than TEM. The focused electron probe of STEM can also be used for device fabrication e.g., to create nanometer sized holes for DNA sequencing[6]. The latter is possible, because most 2D materials are readily damaged by knock-on displacement when the primary electron energy is higher than about 80 keV[7]. Therefore, the development of powerful low-voltage (e.g., 60 keV) aberration-corrected (AC) STEM systems with atomic resolution capabilities, has been indispensable for 2D materials' research. Nowadays, with AC-STEM, atomic resolution imaging of 2D materials with primary electron energies of 60 keV is possible on a routine basis. One advanced TEM has atomic resolution capabilities with ultralow primary electron energy of 20 keV[8]. However, this system uses conventional TEM and therefore does not have the same versatility as a STEM.

Enhancing imaging sensitivity is as important as pushing the imaging resolution at reduced voltages, i.e. resolution is useless without contrast. Emerging techniques such as iDPC-

STEM[9,10] and the recently demonstrated super-resolution electron ptychography[11,12] are capable of reaching higher sensitivity than the conventional annular dark-field (ADF) STEM technique. With this latter technique, relatively light elements are invisible or have reduced contrast, possibly below the noise level, as it images the square of the electrostatic potential[10,13], providing nearly atomic number squared ($Z^{1.6-2.0}$) contrast[14]. On the other hand, iDPC-STEM or electron ptychography are capable of atomic electrostatic potential imaging, which is about linear in $Z$[9,11,12,15]. This substantially improves sensitivity as we have recently demonstrated by the direct imaging of hydrogen atoms in titanium hydride using iDPC-STEM[16].

Observing dynamical phenomena, like phase transitions or single atom motion, is generally more valuable for the fundamental understanding of materials than single snapshots of static materials. However, such experiments require a STEM imaging technique that is not only sensitive, but also has high framerate capabilities, which to date has not been demonstrated. In this respect, iDPC-STEM has a major advantage over electron ptychography, and 4D STEM in general, because the several solid-state electron detectors (in our case four) that iDPC-STEM employs are two to three orders of magnitude faster than electron detecting cameras used for electron ptychography and 4D STEM. For instance, four image frames of 512 x 512 pixels are acquired in about 1 minute with iDPC-STEM in this work (and up to 100 times higher frame rates are possible in general), but take about 35 minutes with 4D STEM, which additionally produce large datasets (68 GB versus 17 MB for iDPC-STEM) and require time consuming and complex reconstructions schemes for electron ptychography[11,12,17].

Here we image defect dynamics in 2D $WS_2$ by recording real-time atomic electrostatic potential movies with iDPC-STEM using a primary electron energy as low as 30 keV. $WS_2$ is a member of the family of transition metal dichalcogenides (TMDs), which have a characteristic layered structure that is easily separable into their monolayer constituents by virtue of the weak intralayer van der Waals forces. The interlayer forces have a mixed covalent-ionic nature and the monolayer consists of a plane of tungsten (W) atoms that is sandwiched between two planes of sulfur (S) atoms. In the $WS_2$ that we study here, the W atoms are trigonal prismatically coordinated (2H structure) with six S atoms, rendering the $WS_2$ monolayer a direct semiconductor. The S atoms overlap when viewed perpendicular to the plane as a result of the trigonal prismatic coordination. The projected atomic number of W and the two S atom stack (which we refer to as $S_2$) is Z=74 and Z=32, respectively. A single sulfur vacancy ($V_S$) has a projected atomic number of Z=16, and a single tungsten vacancy ($V_W$) or a double sulfur vacancy ($V_{2S}$) yields a local vacuum. The relatively large contrast in projected atomic number is the rationale for using here an atomic electrostatic potential imaging technique.

In Fig. 1 we compare simultaneously acquired atomic resolution iDPC-STEM and ADF-STEM images of a 2H $WS_2$ monolayer. The two selected regions (Fig. 1a-d and Fig. 1e-h) contain various elements and defects: W, $S_2$, $V_W$, $V_S$ and $V_{2S}$. All of these features are present in the atomic electrostatic potential images captured using iDPC-STEM, with clearly distinguishable contrast between $S_2$, $V_S$ and $V_{2S}$. The ADF-STEM images also contain signals from both W and $S_2$, however, without appreciable contrast between $S_2$, $V_S$ and $V_{2S}$. The intensity line profiles shown in Fig. 1i,j quantify the critical improvement in sensitivity of iDPC-STEM compared to ADF-STEM to detect all atoms and defects present.

The atomic electric field is displayed as a (colorized) vector field and vector magnitude image in Fig. 1c,d and Fig. 1g,h. These images are two different representations of the same differential phase contrast (DPC) STEM vector image that is complementing the atomic electrostatic potential (iDPC-STEM) scalar image[9,15,18]. In these images, nodes exist at positions where the projected electric field is zero. This occurs at high-symmetry points where the electric field cancels out i.e., at the atom positions, bridge sites and hollow sites. Anomalies such as $V_S$ and $V_{2S}$ are readily detected by their distinct shape in atomic electric field images as outlined in Fig. 1c,d,g,h. Moreover, the sulfur atoms and its vacancies remain detectable even when the $WS_2$ is tilted a few degrees with respect to the electron beam optical axis (supplementary information (SI) Fig. S2). Hence, both atomic electrostatic potentials and atomic electric fields, which are retrieved from the same detector data, have robust single atom sensitivity and are a powerful method to detect light elements.

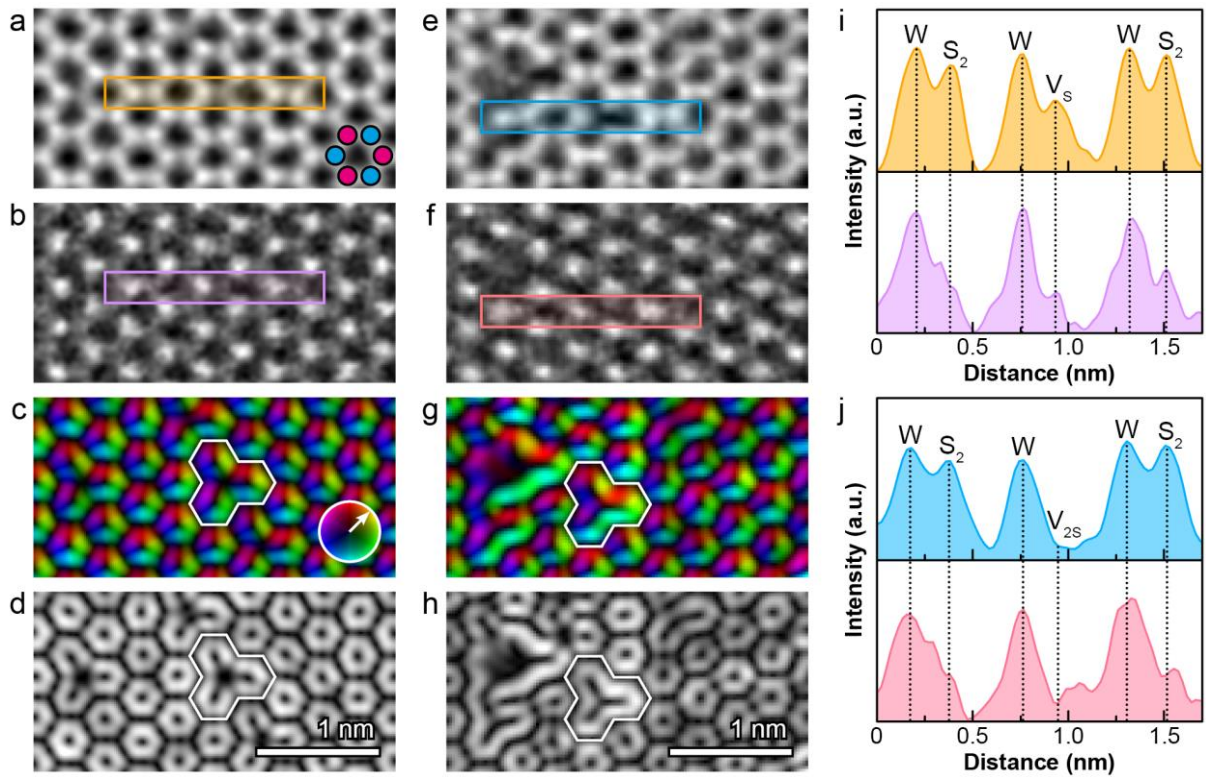

**Figure 1.** Atomically resolved experimental 30 keV-STEM images of a 2H $WS_2$ monolayer with single ($V_S$) and a double ($V_{2S}$) sulfur vacancies. The $WS_2$ schematic in (a) indicates the position of W and $S_2$ with magenta and cyan dots, respectively, and applies to all images. Simultaneously acquired images of two regions with different defects are shown in (a-d) and (e-h). The $V_S$ and $V_{2S}$ defects are indicated in (c,d) and (g,h). Field strength in (c,g) is represented according to the color wheel inset in (c). The images in (a-d) and (e-h) are, from top to bottom: electrostatic potential (iDPC-STEM), square of the electrostatic potential (ADF-STEM), electric vector field (DPC-STEM) and magnitude of electric field (magnitude of DPC-STEM). Line profiles of W, $S_2$, $V_S$ and $V_{2S}$ are extracted from the marked rectangles in the iDPC-STEM (a,e) and ADF-STEM (b,f) images and are plotted in (i,j).

We observe the formation of point defects, defect complexes and agglomerates, one-dimensional (1D) defects and holes under influence of the 30 keV electron beam (Movie S1-S3). To capture these dynamics with a large field of view and a practically useful framerate (e.g., 10-20 seconds per frame of a 10 x 10 nm$^2$ area), a lower effective electron dose than in Fig. 1 has to be used. The lower electron dose (from 1900 to 800 e$^-$/pixel, see SI) typically causes loss of signal from the light elements such as $S_2$ and S in the ADF-STEM images, such that then only contrast from W remains. This is not the case for the atomic electrostatic potential image where all features remain detectable, as has been demonstrated in zeolites[10,19] and metal-organic frameworks[20], due to the intrinsic high sensitivity and noise suppression property of iDPC-STEM[9,10].

We first show the creation and dynamics of $V_W$, $V_S$ and $V_{2S}$ point defects by four subsequent frames in Fig. 2. The defects present are schematically indicated in the electric field magnitude images Fig. 2e-h. The time interval between the frames is in this case 14.0 seconds, but these frames were cut from a larger overview and the actual time to record each field of view shown was 1.0 second. In the first frame, $V_S$ and $V_{2S}$ are randomly distributed, which agglomerate in the second frame. Hence, the point defects are mobilized under influence of the 30 keV electron beam. The defects are particularly visible in the electric field and atomic electrostatic potential images but are practically invisible in the ADF-STEM image. The third frame shows that a W atom, originally residing at the center of the sulfur vacancy agglomerate, has moved about 1 nm to a W-W bridge position. The ejected W atom was likely destabilized due to undercoordination at its original position in the sulfur defect agglomerate. Finally, in the fourth frame the W atom has moved back to its original site and sulfur defects have lined up to create a 1D sulfur vacancy line (SVL)[21,22]. Note that this process is largely invisible in the ADF-STEM image, except for the short moment (single frame) the tungsten vacancy is present.

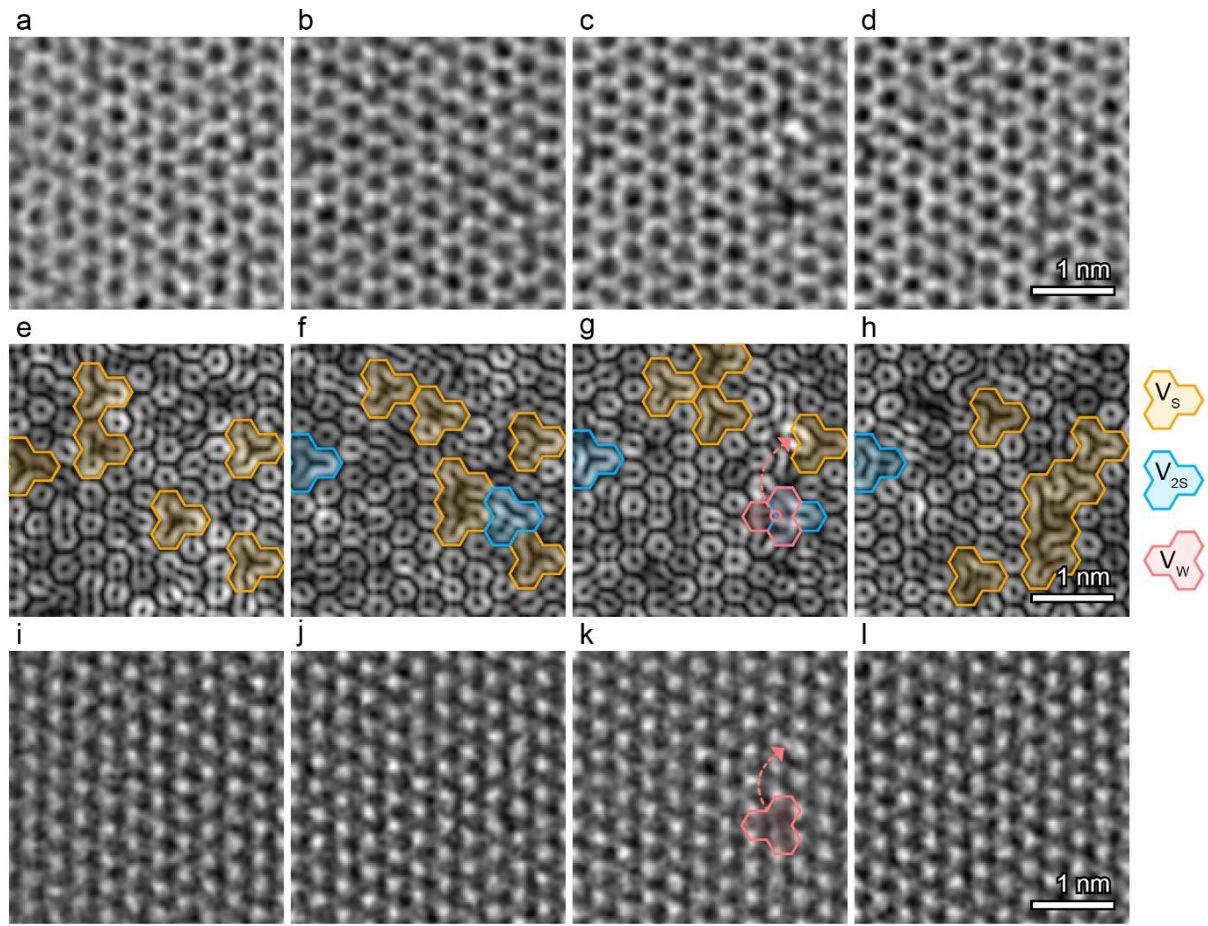

**Figure 2.** Four frame sequence displaying point defect creation and motion induced by the 30 keV electron beam. The rows display simultaneously acquired images of the electrostatic potential (iDPC-STEM) (a-d), electric field magnitude (DPC-STEM) (e-h), and square of the electrostatic potential (ADF-STEM) (i-l). The defects present are schematically indicated in the electric field magnitude images.

We observed that formation of SVLs is the dominant mechanism to accommodate sulfur vacancies in $WS_2$ upon exposure to the 30 keV electron beam, similar to what has been observed for $MoS_2$ with an electron beam energy of 60 keV or when the specimen is heated to elevated temperatures[23]. An example of this process is shown in Fig. 3, where the same area as in Fig. 2 is presented, but 15 frames later.

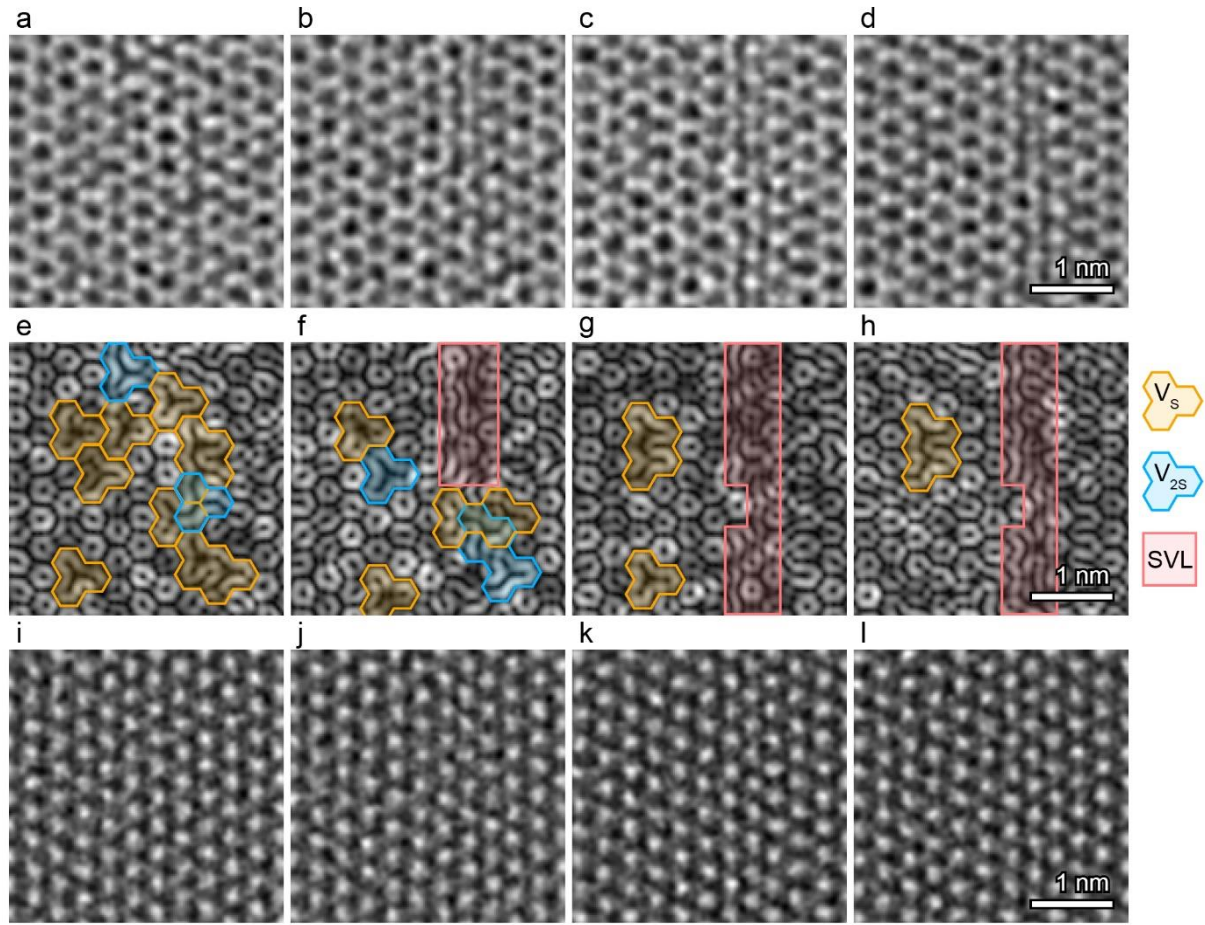

**Figure 3.** Four frame sequence showing how point defects agglomerate into line defects induced by the 30 keV electron beam. The columns display simultaneously acquired images of the electrostatic potential (iDPC-STEM) (a-d), electric field magnitude (DPC-STEM) (e-h), and square of the electrostatic potential (ADF-STEM) (i-l). The defects present are schematically indicated in the electric field magnitude images.

In Fig. 3a,e the defect density is high, about 1.4 $V_S/nm^2$ (i.e., locally $1.4 \times 10^{14}$ $V_S/cm^2$). The highly defective $WS_2$ crystal appears to be deformed through local buckling or local reconstruction that deviates from a perfect crystal[23]. We infer this from the apparent lower quality of the image, which is caused by the smeared-out intensity of $S_2$ atomic electrostatic potentials. The separate atoms in misaligned $S_2$ cannot be resolved as the projected distance between them is below the attainable resolution. Hence, misaligned $S_2$ form a single feature with reduced intensity (approaching the one of a single S atom) and smeared-out appearance. In most cases it is possible to differentiate between $V_S$ and misaligned $S_2$, because $V_S$ is more spatially focused than the smeared-out misaligned $S_2$. Upon further exposure, the majority of $V_S$ concentrate and form a short SVL. The SVL is two rows of $V_S$ wide, and is observable in the top right part of the electric field image as well as the atomic electrostatic potential image (Fig. 3b,f). Note that the SVL is only indirectly visible in the ADF-STEM image, shown in Fig. 3j, by a decreased nominal W-W distance across the SVL. Another SVL has nucleated and nearly merged with the existing SVL (Fig. 3c,g). Only in the center two more $V_S$ are required to merge the SVL over its full width. The final frame shows how the overall crystallinity of $WS_2$ has improved compared to the initial state, by accommodating sulfur vacancies in SVLs.

Holes are created as well in $WS_2$ after prolonged exposure (nearly 50 frames after Fig. 3) to the 30 keV electron beam. Prior to this, the $WS_2$ is in a highly defective state with a high density of point defects and SVLs. Further exposure first leads to loss of W atoms and is then (within several frames) followed by rapidly growing holes. We observe single atom motion at the exposed hole edges, of which the dynamics at W-terminated zigzag edges are displayed in Fig. 4. The two edges are both terminated by W atoms, but they have a different symmetry, because one of the edges is a true W-terminated zigzag edge (diagonal edge), whereas the other edge (vertical edge) was originally an S-terminated edge, where the S atoms have been removed by the electron beam. Here we show how W and S atoms are appearing at the edges and bond to exposed undercoordinated W edge atoms. This is evident beyond any doubt in the atomic electrostatic potential images, and just noticeable in the electric field images as well, although more difficult to interpret if they would have to stand alone. In contrast, in the ADF-STEM image only the W atom is visible, and a lot of atomic structure details are thus missing.

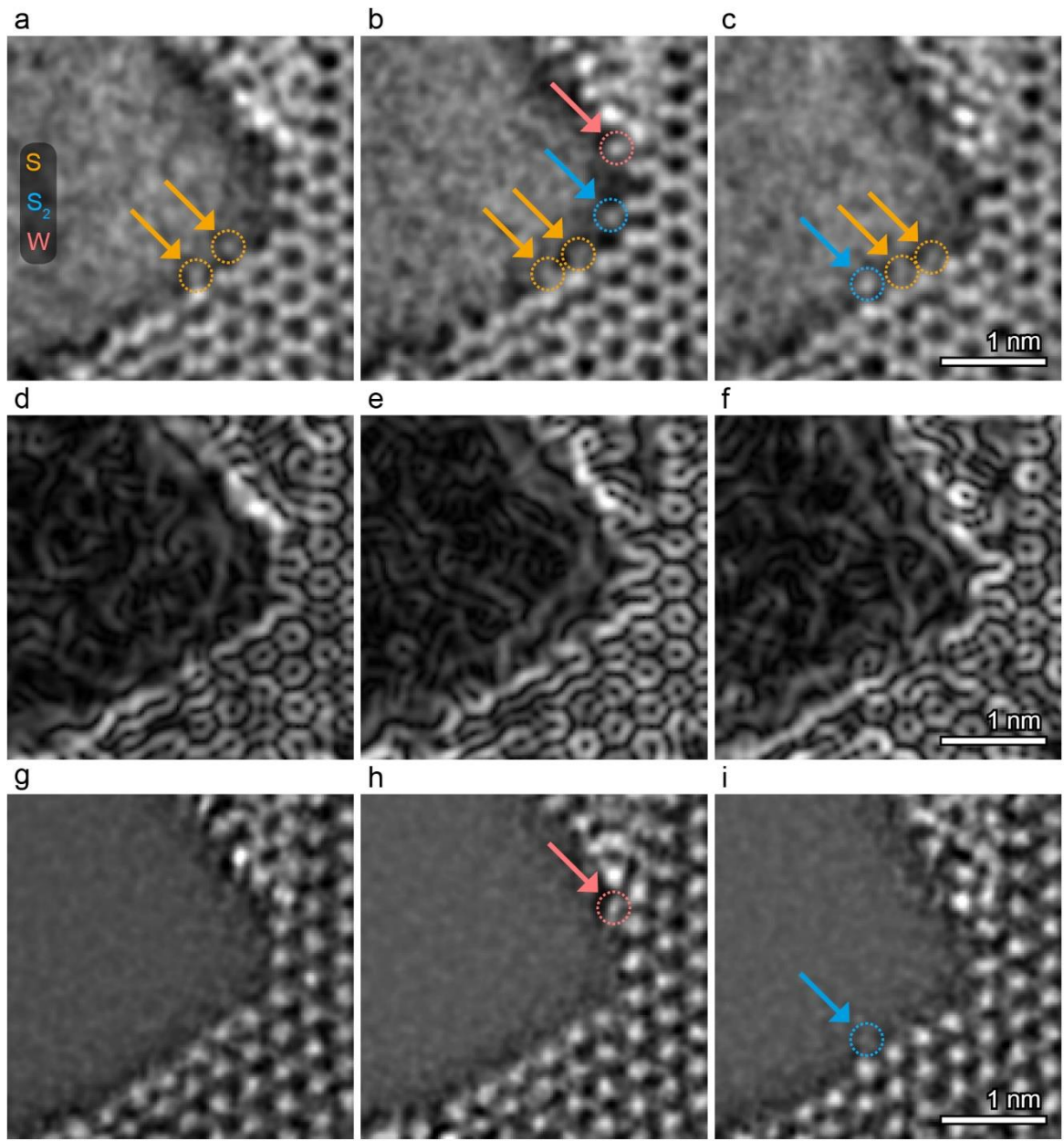

**Figure 4.** Sequence of frames revealing single atom motion at $WS_2$ edges induced by the 30 keV electron beam. The rows display simultaneously acquired images of the atomic electrostatic potential (iDPC-STEM) (a-c), atomic electric field magnitude (DPC-STEM) (d-f), and square of the electrostatic potential (ADF-STEM) (g-i). Colored arrows point at specific edge atoms present.

Attaining atomic resolution with 30 keV electrons is fundamentally more challenging than using more conservative energies of 60 keV and 80 keV, because the resolution is proportionally limited by the electron wavelength, which, at 30 keV, is longer by about 43% and 67% compared to 60 and 80 keV, respectively. To achieve atomic resolution with the longer wavelength, advanced and highly-stable geometrical aberration correctors are required that enable the increase of the numerical aperture of the probe forming lens (see SI). In addition, a more monochromatic electron beam is necessary to minimize probe broadening due to chromatic aberrations. This can be achieved with a cold field-emission electron gun or an electron monochromator as we use here, which limits the available electron dose and thus demands the use of a dose efficient imaging technique such as iDPC-STEM.

The strong efforts in the last decade to achieve atomic resolution at accelerating voltages much lower than the traditional 200 and 300 kV were strongly motivated by prospects of reduced electron beam damage allowing materials to be studied in their intrinsic state for prolonged times. Indeed, the knock-on displacement damage is greatly reduced when lowering the accelerating voltage to 80 or 60 kV. In particular for carbon-based materials, like carbon nanotubes and graphene, this development towards lower accelerating voltages has been very successful. However, caution is required since damage by other mechanisms, in particular radiolysis, aggravate for less conductive materials such as semiconductors and insulators. The present results show that the 2D $WS_2$ analyzed here is not inert to the 30 keV electron beam. Sulfur vacancies are readily created, and they diffuse through the material as is evident from the formation of SVLs. The mechanisms that provide energy for such processes must be dominantly related to ionization effects and to a much lesser extent knock-on displacement. In particular, the ionization cross-section increases with decreasing electron beam energy and is more profound in semiconducting materials (which is the case here for the 2H $WS_2$)[24,25]. On the other hand, 30 keV is well below the knock-on displacement threshold of S and W, since, respectively, only about 2.1 eV and 370 meV is transferred upon head-on collision of the primary 30 keV electron beam with the atom nuclei. A comparison with 60 keV for the same WS$_2$ material shows that the mechanism and dynamics of damage formation are rather different for 30 and 60 keV, even worsening the situation at the lower accelerating voltage. Results of this comparison will be presented in a follow-up paper. More in general, the electron beam damage mechanisms at play at these low beam energies are not fully understood, and remain an active area of research[26,27].

The ability to image atomic electrostatic potentials of 2D WS$_2$ with high framerate at 30 keV is an improvement over the previously reported 30 keV ADF-STEM image of graphene[28]. Mainly due to the fact that graphene is not susceptible to beam damage at these low voltages, such image quality can be improved by increasing the electron dose. The 2D WS$_2$ that we study here, however, is sensitive to ionization damage and directly restricts the electron dose. We have demonstrated that the image quality of ADF-STEM deteriorates under these conditions, whereas atomic electrostatic potential imaging remains of high quality due to the inherent high sensitivity and noise suppression property of iDPC-STEM.

Consequently, capturing the atomic electrostatic potential image with high framerate is not only possible from an image quality perspective, but also from a technical perspective, because iDPC-STEM employs solid-state electron detectors that have equivalent readout speeds to, for example, ADF detectors. In the iDPC-STEM images, however, the contrast between W and S$_2$ is limited, such that the combination of the ADF-STEM and iDPC-STEM image is still necessary to differentiate between them. There are two reasons that contribute to this effect. The dominant contribution is of fundamental nature: the finite resolution of the electron microscope bandwidth limits the potential from W atoms more strongly than the S atoms[29], reducing the contrast of $I_W/I_{S2}$ from about 1.8 to 1.25. The other smaller contribution is the finite accuracy of iDPC-STEM to image atomic electrostatic potentials, which increases the relative intensity of S compared to W (see SI and Fig. S3).

Atomic electrostatic potential imaging can be achieved with absolute accuracy with electron ptychography, or when truly integrated center-of-mass (iCOM) STEM imaging is performed[9,10], which is only approximated (although rather well with even only four detector segments) by iDPC-STEM. The advantage of electron ptychography is that it can further improve contrast with its super-resolution capabilities. However, iDPC-STEM is considerably more suitable for dynamical phenomena studies than electron ptychography, due to its direct imaging and vast (orders of magnitude) speed advantage. Moreover, the accuracy of iDPC-STEM can be readily improved, approaching the absolute accuracy of iCOM-STEM, by employing more solid-state electron detectors than the four that we use here, without compromising speed.

Furthermore, simultaneous imaging and electron energy loss spectroscopy (EELS) cannot be performed with electron ptychography, because all electrons are blocked by the camera, although recently the idea of performing electron ptychography using a camera with a central hole has been reported[30]. iDPC-STEM is directly compatible with EELS, as its detector already has a central hole. This is of particular importance considering the advent of ultra-high energy resolution EELS[31], and fast and sensitive direct electron detectors for EELS[32] that are opening up new fields in electron microscopy e.g., the ability to directly measure vibrational spectra to investigate phonon modes in nanostructured materials at atomic resolution[33–36].

In summary, we have demonstrated that defect dynamics in 2D $WS_2$ can be imaged with enhanced sensitivity and high framerates using the ultralow electron beam energy of 30 keV.

This has been enabled by the atomic electrostatic potential imaging capability of iDPC-STEM and advancements in electron optics. At present, this is the only STEM imaging technique that combines high sensitivity and high framerates. The real-time iDPC-STEM movies reveal light sulfur atoms and their dynamics in $WS_2$ that are invisible with the traditional ADF-STEM. This approach can be directly applied to visualize light elements, like oxygen, carbon and nitrogen, in all 2D materials, and can be generalized to all other beam sensitive materials that require low electron dose and ultralow beam energies. The possibility of combining this fast and sensitive imaging technique with powerful emerging electron spectroscopic capabilities has the potential to solve challenging problems in materials science.

**Supporting Information**

The Supporting Information is available free of charge.

Sample preparation of 2D WS2 for scanning transmission electron microscopy analysis; Scanning transmission electron microscopy experimental conditions and image processing; Robust electrostatic potential imaging of light elements in tilted 2D WS2; Simulated contrast and accuracy of atomic electrostatic potential imaging of 2D WS2 with iDPC-STEM (PDF).

Movie S1 (AVI).

Movie S2 (AVI).

Movie S3 (AVI).

**Corresponding Authors**

*sytze.de.graaf@rug.nl, b.j.kooi@rug.nl

**Author Contributions**

The manuscript was written through contributions of all authors. All authors have given approval to the final version of the manuscript

**Funding Sources**

Financial support from the Zernike Institute for Advanced Materials and the Groningen Cognitive Systems and Materials Center is gratefully acknowledged.

**Acknowledgements**

T.S. Ghiasi, J. Peiro, A. Kaverzin are thanked for their support in TEM specimen preparation. M. Liang is acknowledged for the CVD grown $WS_2$.

**Supplementary information for**

# Real-time imaging of atomic potentials in 2D materials with 30 keV electrons

*Sytze de Graaf[1,\*], Majid Ahmadi[1], Ivan Lazić[2], Eric G.T. Bosch[2], Bart J. Kooi[1,\*]*

[1]Zernike Institute for Advanced Materials, University of Groningen, Nijenborgh 4, 9747 AG Groningen, Netherlands

[2]Thermo Fisher Scientific, Achtseweg Noord 5, 5651 GG Eindhoven, Netherlands.

**Sample preparation of 2D WS$_2$ for scanning transmission electron microscopy analysis**

Monolayer flakes of WS$_2$ were either transferred to, or directly grown on, Si/SiO$_2$ (300 nm) substrates by, respectively, exfoliating bulk crystals (HQ Graphene) with the scotch tape method for Fig. 1a-d and Fig. 2-4, or using chemical vapor deposition (CVD) for Fig. 1e-h. The monolayer flakes were identified using optical microscopy and transferred to R 1.2/1.3 carbon Quantifoil TEM grids with a polymer free method, as shown in Fig. S1.

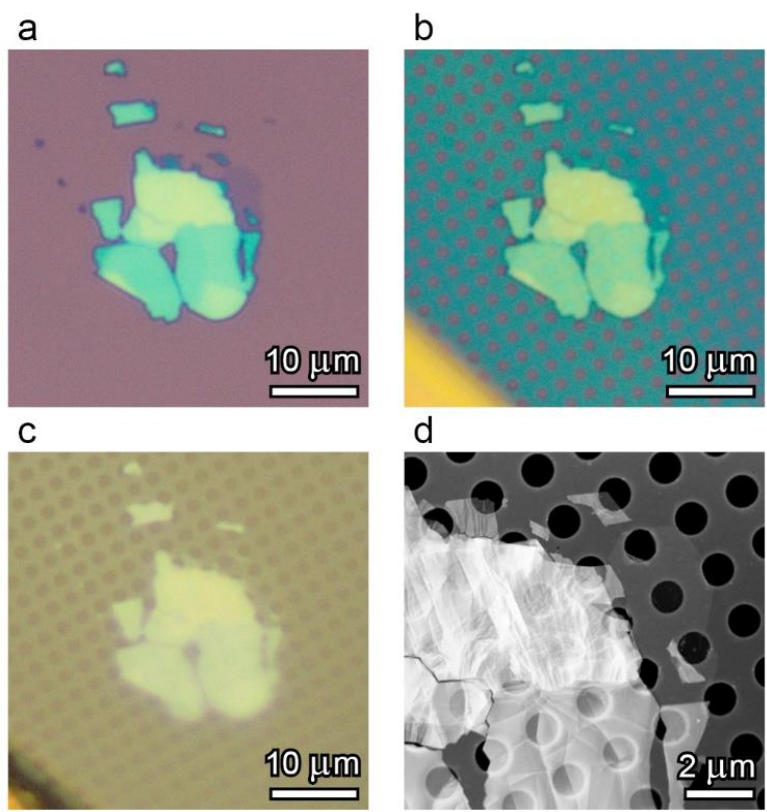

**Figure S1.** Microscope images of the exfoliated WS$_2$ flake taken after different steps of the transfer method. (a) The exfoliated WS$_2$ flake on Si/SiO$_2$ (300 nm) substrate, with a monolayer part in the top right (dark contrast). (b) Perforated carbon support grid of the TEM grid is attached to the flake after the isopropanol has evaporated. Bottom left shows a part of the gold grid. (c) TEM grid with WS$_2$ flake is detached from the Si/SiO$_2$ substrate after etching in 1M KOH. (d) Overview ADF-STEM image of the WS$_2$ flake. Weak contrast of the monolayer part is visible at the upper right part of the flake.

First, the TEM grid is carefully put on top of the monolayer $WS_2$ with the carbon support film facing the flake. One droplet of isopropanol is placed on the TEM grid, which directly wets both the carbon film and the $SiO_2$ substrate to form a liquid interface between them. The flexible TEM grid is pulled towards the flake and the substrate upon evaporation of the isopropanol, which results in strong adhesion of the TEM grid to the $WS_2$ flake and substrate. The entire stack (substrate/flake/TEM grid) is then immersed in a 1M KOH solution at room temperature to etch away the $SiO_2$, for as long as necessary to detach the TEM grid (usually 5-15 minutes). Then the TEM grid is carefully extracted from the solution with self-closing tweezers, which are then dried together (as otherwise the TEM grid sticks to the tweezer upon opening due to the liquid capillary) for 5-10 minutes in an atmospheric furnace at 70-80 °C. The dried TEM grid is floated on deionized water (with carbon support film facing up) for 15-30 minutes to rinse off precipitated KOH crystals. Note that the drying step prior to rinsing in water is crucial, as otherwise the carbon film is found to very easily detach from the gold grid when floating the grid on water. Finally, the grid is scooped with filter paper and dried for 5-10 minutes in an atmospheric furnace at 70-80 °C.

**Scanning transmission electron microscopy experimental conditions and image processing**

Here we have used a state-of-the-art 30-300 kV Thermo Fisher Themis Z scanning transmission electron microscope. The microscope is equipped with a high coherence, high brightness field-emission electron gun (X-FEG), electron monochromator, probe $C_S$ corrector (S-CORR), image $C_S$ corrector (CETCOR) and a segmented DF4 detector for iDPC-STEM imaging.

The microscope was stabilized over the weekend after changing the high-tension to 30 kV. Then the aberrations up to the 5$^{th}$ order were minimized with the probe corrector software to the measured values shown in Table 1. During the experiments, the first and second order aberrations were minimized using the OptiSTEM software, which employs a contrast optimization algorithm of atomic resolution ADF-STEM images. Finetuning of defocus (C1), first order astigmatism (A1) and coma (B2) was done manually.

These low aberration values produce a phase front with a phase difference less than $\pi/2$ up to 50 mrad, such that an optimal small electron probe can be obtained with a convergence semi-angle up to this value. In our case, however, the largest suitable limiting aperture was 33 mrad, restricting the fundamental achievable resolution to 1.06 Å, which we closely reached in experiment. Because we do observe the WS$_2$ (20-20) planes (1.36 Å) and not the next following (12-30) planes (1.03 Å). Resolution below 1 Å can be obtained with larger better matching limiting apertures such that the probe corrector is used to its full potential.

**Table 1.** A list of measured aberration values and confidence levels after tuning the probe $C_S$ corrector at 30 keV.

| Aberration | Value | 95% confidence |
|---|---|---|
| WD | 300 μrad | N/A |
| C1 | 3 nm | 1 nm |
| A1 | 2 nm | 1 nm |
| A2 | 13 nm | 43 nm |
| B2 | 28 nm | 42 nm |
| C3 | -540 nm | 1 μm |
| A3 | 210 nm | 400 nm |
| S3 | 220 nm | 330 nm |
| A4 | 3.4 μm | 2.7 μm |
| D4 | 2.6 μm | 1.8 μm |
| B4 | 4.2 μm | 3.4 μm |
| C5 | 240 μm | 320 μm |
| A5 | 240 μm | 47 μm |
| S5 | 100 μm | 74 μm |

At these ultra-low electron beam energies, the probe broadening due to chromatic aberrations has substantial impact. For instance, atomic resolution could hardly be achieved without reducing the energy spread of the electron source. Hence, the energy spread was reduced by employing the monochromator, which was excited to a value of 0.35, after which the microscope was stabilized overnight. With the circular energy selecting apertures of 0.5 and 1.0 µm that we employed, the maximum available probe current was about 3 pA and 8 pA, respectively. The collection angles of the segmented DF4 detector and ADF detector were 9-36 mrad and 39-200 mrad, respectively.

For Fig. 1 in the main article an image of 512x512 pixels was acquired with 7.5 pA probe current, 40 µs pixel dwell time and a step size of 15.92 pm. For Figs. 2-4 in the main article, a 200-frame movie of 512x512 pixels was acquired with 2.5 pA probe current, 50 µs pixel dwell time and a step size of 22.52 pm. Thus, the electron doses were about 1900 e$^-$/pixel and 800 e$^-$/pixel for Fig. 1 and Figs. 2-4, respectively. Despite the larger energy spread, due to using a larger energy selecting aperture, the image quality of Fig. 1 is predominantly improved as a result of the more than twice as high electron dose. Also, the smaller step size used contributed to the improved image quality.

Specimen drift is another critical component that should be minimized in order to be able to continuously capture the same field of view for prolonged acquisition times. The original images, from which Figs. 2-4 in the main article were extracted, have a field of view of about 11 x 11 nm$^2$, and were acquired with a frame rate of 14.0 seconds per frame, yielding a total acquisition time of 2800 seconds, or about 45 minutes. In this case the specimen drifted 2.5 nm within the first 50 frames yielding a drift rate of 0.22 nm/min. Thereafter, in the remaining 150 frames, the specimen was virtually stable with a total drift distance of 1 nm (drift rate of 0.03 nm/min). This is possible by the active vibrational damping system of the microscope column, ultra-stable room temperature (maximal deviation of 0.2 °C/24 hours), and the use of a piezo controlled stage.

The ADF- and iDPC-STEM images were bandpass filtered with Fiji[1,2] using the 'FFT Bandpass Filter' with a low-pass filter of 3 pixels, and a high-pass filter of 10 pixels. The DPC-STEM images were bandpass filtered in MATLAB with a Gaussian low-pass filter with standard deviation of 3.0 nm$^{-1}$, and a Gaussian high-pass filter with a standard deviation of 1.0 nm$^{-1}$. We explicitly note that low-pass filtering can be safely applied to any STEM imaging technique, as only noise and no information is present at frequencies beyond twice the convergence semi-angle (see below for detailed explanation). We observe that particularly the quality of ADF-STEM images improves by low-pass filtering, as a substantial amount of high-frequency noise is present that otherwise totally obscures the light elements in the raw images. On the other hand, high-pass filtering is beneficial for iDPC-STEM images. Because iDPC-STEM has strong contrast transfer at low-frequency information, which is mainly surface adsorbates and carbon contamination in this case. We refer the reader to our previous work for a comparison of filtering on ADF- and iDPC-STEM images[3]. Furthermore, the DPC-STEM images were deliberately strongly low-pass filtered (removing not only high-frequency noise but also high-frequency information) to enhance visibility of defect features.

The separate frames of Fig. 2-4 are compiled to a movie and are attached as Movie S1, Movie S2 and Movie S3, respectively. There simultaneously acquired iDPC-STEM, DPC-STEM and ADF-STEM images are shown, with the effective elapsed time (for the field of view) indicated in the top left corner.

**Robust electrostatic potential imaging of light elements in tilted 2D WS$_2$**

The capability of robust imaging of misaligned materials is particularly valuable in the case of transition metal dichalcogenides (TMDs) with the 2H symmetry. Because in that case the two chalcogen atoms in a monolayer overlap in projection when it is perfectly aligned along the (0001) crystal zone axis, boosting its atomic resolution signal. However, in practice the 2D TMDs can be easily misaligned with respect to the optical axis of the electron beam. Slight bending/buckling of the suspended 2D material flake on the TEM grid can be expected, e.g. as a result of the strong capillary forces of the liquid used during TEM sample preparation that create residual stress upon drying. Hence, it is a challenging task to well-align the 2D material locally, since the alignment fluctuates on a lateral scale of just tens of nanometers. In Fig. S2 we show the simultaneously acquired iDPC-STEM and ADF-STEM image of a (few degrees) misaligned monolayer 2H WS$_2$. These results demonstrate that it is still possible to properly detect and distinguish single and double sulfur vacancies ($V_S$ and $V_{2S}$) with iDPC-STEM, but not with ADF-STEM, which nearly lost all its sensitivity towards sulfur.

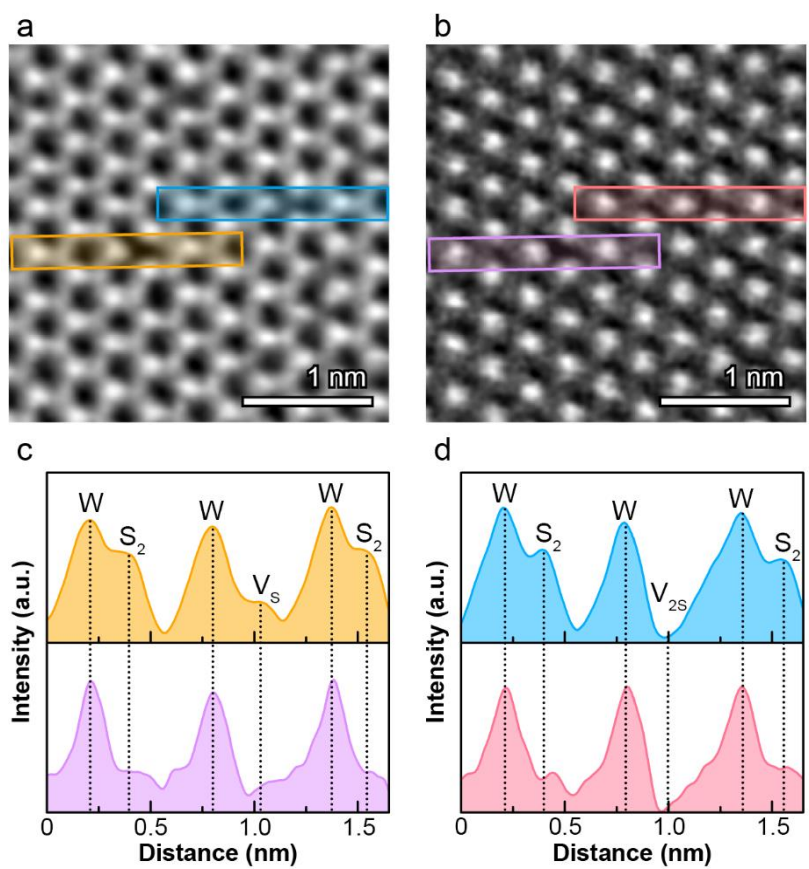

**Figure S2.** Images of a monolayer 2H $WS_2$ with $V_S$ and $V_{2S}$ were acquired from a region that was misaligned by a few degrees from the (0001) zone axis. It shows how iDPC-STEM (a) robustly images the sulfur vacancies with single sulfur atom sensitivity as opposed to ADF-STEM in (b). Line profiles from $V_S$ and $V_{2S}$ are shown in (c,d), from the rectangular regions marked in (a,b).

**Simulated contrast and accuracy of atomic electrostatic potential imaging of 2D $WS_2$ with iDPC-STEM**

The finite resolution of the electron microscope poses a fundamental limit on the contrast between W and $S_2$ in the case of a 2H $WS_2$ monolayer. Specifically, in the case of ADF-STEM and iDPC-STEM, information is only transferred up to a wavevector (i.e., reciprocal length or frequency) that is twice the length of the largest wavevector present in the electron probe. In our case, with a convergence semi-angle of 33 mrad and electron wavelength of about 7 pm, the largest wavevector present in the electron probe is 4.7 $nm^{-1}$. Which corresponds to maximum information transfer at 9.4 $nm^{-1}$, or 1.06 Å.

In Fig. S3 we have simulated the atomic potential of a 2H $WS_2$ monolayer using the multislice approach with the Dr. Probe software[4]. We neglect atomic thermal vibrations and we assume an aberration free electron probe, to demonstrate the effects of finite resolution and reduced accuracy of atomic potential imaging with iDPC-STEM. In Fig. S3a the simulated atomic potentials of a 2H $WS_2$ monolayer are shown, which we use as the 'true ground state' for further comparison. The bandwidth limited atomic potentials of $WS_2$ are shown in Fig. S3b. The corresponding line profiles in Fig. S3e,f demonstrate how the contrast between W and $S_2$ reduces from a relative intensity of 1.8 to 1.25 by imposing the fundamental information transfer limit. However, this reduction would be somewhat less when atomic vibrations are considered. Because in that case also the 'ground state' atomic potentials already would have reduced contrast between W and $S_2$.

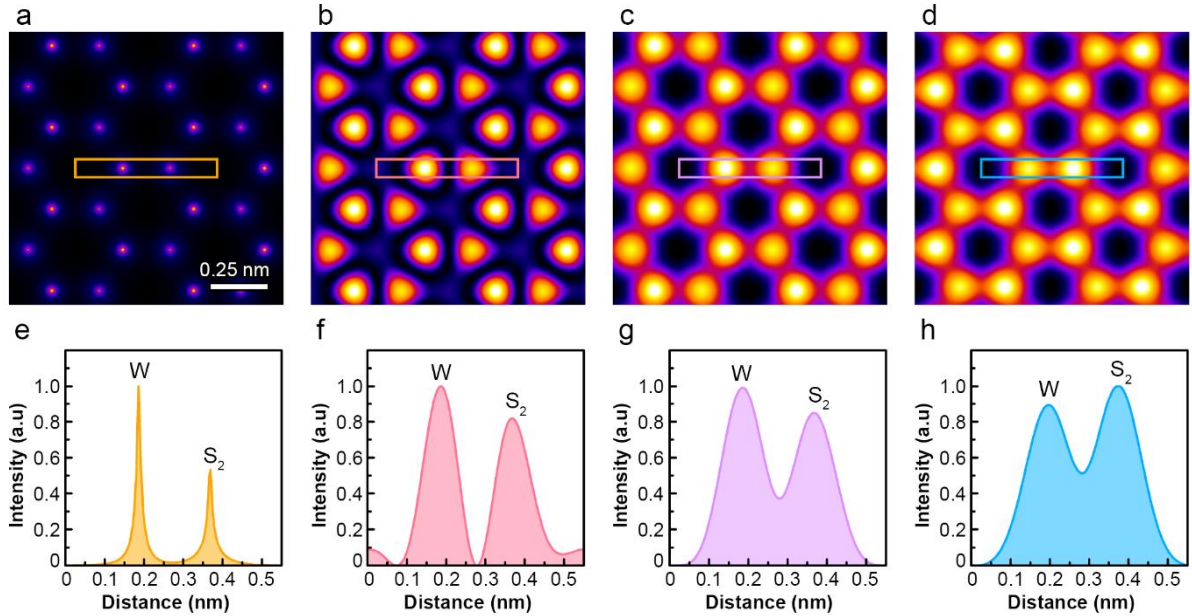

**Figure S3.** Simulated images of atomic electrostatic potentials of 2H WS$_2$ monolayer using the multislice approach. (a) The 'true ground state' atomic electrostatic potentials, which are the input for multislice simulations. (b) The atomic electrostatic potentials from (a) are bandwidth limited to the maximum transferrable wavevector to obtain the highest possible resolution image. (c) A simulated iCOM-STEM image (ideal version of iDPC-STEM when enough segments are used) based on the atomic electrostatic potentials in (a). (d) A simulated iDPC-STEM image using four detector segments based on the atomic electrostatic potentials in (a). Line profiles of the W and S$_2$ atomic electrostatic potentials are plotted in (e-h), corresponding to the marked rectangular area in (a-d).

In fact, the contrast is lower in the case of iCOM-STEM (Fig. S3c,g) and iDPC-STEM (Fig. S3d,h), because then less high frequency information transfer occurs than that assumed in Fig. S3b. The reason is that the contrast transfer functions (CTF) of iCOM-STEM and iDPC-STEM decreases linearly to zero at the maximum wave vector, whereas the (imposed) contrast transfer function for Fig. S3b keeps a value of one up to the maximum wave vector and then in a step becomes zero.

Note that iCOM-STEM can be considered the most ideal case of iDPC-STEM, that can be achieved when enough detector segments are used. It is important to stress here that the object of imaging (unlike in ADF-, (A)BF- or DPC-STEM) is truly the projected atomic electrostatic potential of the sample, rendering iCOM-STEM (without any approximations) a linear imaging technique. The CTF of iCOM-STEM is positive definite (no zero crossings) and with full range frequency transfer near the optimal focus. With larger segment detectors, like commonly used in iDPC-STEM, the COM is determined with less precision (although even with only four detector segments rather well). Nevertheless, even in that case, iDPC-STEM is a dominantly linear imaging technique, with well-defined CTF (reflecting the symmetry of the detector used) including almost negligible correction terms[5]. Contribution of these terms become noticeable only far out of focus for thin samples or in general for rather thick samples[6], where part of the sample is always out of focus.

The iDPC-STEM image (Fig. S3d,h) reveals that the intensity of the $S_2$ peak even exceeds that of W. This is a direct result of using only four detector segments to approximate the COM, but the accuracy can be readily improved by employing more detector segments. In experiment, however, we observe that the heavier W atom are usually equally bright or slightly brighter than the $S_2$ atoms. This deviation between simulated and experimental iDPC-STEM images is probably related to neglecting chromatic aberration in the simulation.